\documentclass[a4paper,fleqn,usenatbib,useAMS]{mnras}

\usepackage{graphicx}	% Including figure files
\usepackage{amsmath}	% Advanced maths commands
\usepackage{amssymb}	% Extra maths symbols
\usepackage{multicol}        % Multi-column entries in tables
\usepackage{bm}		% Bold maths symbols, including upright Greek
\usepackage{pdflscape}	% Landscape pages
\usepackage{multirow}
\usepackage{xcolor}
\usepackage{hyperref}
\usepackage{booktabs}
\usepackage{subfigure}
\usepackage{makecell}
\usepackage{tikz}

\newcommand{\kms}{\,km\,s$^{-1}$} % kilometres per second
\def\specialname[#1]{\textbf{\textsc{#1}}}

\definecolor{lime}{HTML}{A6CE39}
\DeclareRobustCommand{\orcidicon}{%
	\begin{tikzpicture}
	\draw[lime, fill=lime] (0,0) 
	circle [radius=0.16] 
	node[white] {{\fontfamily{qag}\selectfont \tiny ID}};
	\draw[white, fill=white] (-0.0625,0.095) 
	circle [radius=0.007];
	\end{tikzpicture}
	\hspace{-2mm}
}
\foreach \x in {A, ..., Z}{%
	\expandafter\xdef\csname orcid\x\endcsname{\noexpand\href{https://orcid.org/\csname orcidauthor\x\endcsname}{\noexpand\orcidicon}}
}

\usepackage[T1]{fontenc}
\usepackage{ae,aecompl}

\usepackage{newtxtext,newtxmath}
\title[Late-formed halos prefer to host quiescent central galaxies]{
    Late-formed halos prefer to host quiescent central galaxies. I.
Observational results }
\author[Wang et al.]{Kai Wang\orcidK{},$^{1}$\thanks{Contact e-mail:
    wkcosmology@gmail.com} Yangyao Chen\orcidC{},$^{2, 3}$ Qingyang Li\orcidL{},$^{4, 5, 6}$ Xiaohu Yang\orcidY{},$^{4,5, 7, 8}$ \\
    $^1$Kavli Institute for Astronomy and Astrophysics, Peking University,
    Beijing 100871, China\\
    $^2$School of Astronomy and Space Science, University of Science and Technology of China, Hefei, Anhui 230026, China\\
    $^3$Key Laboratory for Research in Galaxies and Cosmology, Department of Astronomy, University of Science and Technology of China, Hefei, Anhui 230026, China \\
    $^4$Department of Astronomy, School of Physics and Astronomy, Shanghai
    Jiao Tong University, Shanghai 200240, China\\
    $^5$Shanghai Key Laboratory for Particle Physics and Cosmology, Shanghai
Jiao Tong University, Shanghai 200240, China\\
    $^6$Institute for Astronomy, University of Edinburgh, Royal Observatory, Edinburgh EH9 3HJ, United Kingdom\\
    $^7$Tsung-Dao Lee Institute, Shanghai Jiao Tong University, Shanghai, 200240, China\\
    $^8$Key Laboratory for Particle Physics, Astrophysics and Cosmology, Ministry of Education, Shanghai Jiao Tong University, Shanghai 200240, China\\
}

\date{Last updated 2020 May 22; in original form 2018 September 5}

\pubyear{2020}

\begin{document}
	\label{firstpage}
	\pagerange{\pageref{firstpage}--\pageref{lastpage}}
	\maketitle

% Abstract of the paper

\begin{abstract}
    The star formation and quenching of central galaxies are regulated by the
    assembly histories of their host halos. In this work, we use the central
    stellar mass to halo mass ratio as a proxy of halo formation time, and we
    devise three different models, from the physical hydrodynamical simulation
    to the empirical statistical model, to demonstrate its robustness. With
    this proxy, we inferred the dependence of the central galaxy properties on
    the formation time of their host halos using the SDSS main galaxy sample,
    where central galaxies are identified with the halo-based group finder. We
    found that central galaxies living in late-formed halos have higher
    quiescent fractions and lower spiral fractions than their early-formed
    counterparts by $\lesssim 8\%$. Finally, we demonstrate that the group
    finding algorithm has a negligible impact on our results.
\end{abstract}

% Select between one and six entries from the list of approved keywords.
% Don't make up new ones.

	%%%%%%%%%%%%%%%%%%%%%%%%%%%%%%%%%%%%%%%%%%%%%%%%%%

	%%%%%%%%%%%%%%%%% BODY OF PAPER %%%%%%%%%%%%%%%%%%

\section{Introduction}%
\label{sec:introduction}

In the concordance $\Lambda$CDM model, the basic blocks of the cosmic structure
are dark matter halos, the virialized parts of the cosmic density field
assembled through gravitational instability. The abundance, spatial
distribution and formation histories of the dark matter halo population are now
well-understood using semi-analytical methods and numerical simulations
\citep[e.g.][]{moAnalyticModelSpatial1996, navarroUniversalDensityProfile1997,
    jingDensityProfileEquilibrium2000, wechslerConcentrationsDarkHalos2002,
    gaoAgeDependenceHalo2005, moGalaxyFormationEvolution2010,
chenRelatingStructureDark2020}. This provides a solid foundation to understand
the galaxy population, as galaxies are believed to form and evolve in dark
matter halos and the properties of galaxies are expected to be closely related
to their host halos \citep{whiteCoreCondensationHeavy1978}.

Many methods have been used to establish the connection between galaxies and
dark matter halos in the framework provided by current cosmology. These include
direct numerical simulations that are based on the first principles of physical
processes combined with some subgrid recipes for modeling unresolved processes
\citep[e.g.][]{vogelsbergerModelCosmologicalSimulations2013,
    schayeEAGLEProjectSimulating2015, daveMUFASAGalaxyFormation2016,
    weinbergerSimulatingGalaxyFormation2017,
nelsonIllustrisTNGSimulationsPublic2019, daveSIMBACosmologicalSimulations2019},
semi-analytical models that are based on a set of empirical functions to
approximate physical processes of galaxy formation and evolution
\citep[e.g.][]{whiteGalaxyFormationHierarchical1991,
    kauffmannFormationEvolutionGalaxies1993,
    somervilleSemianalyticModellingGalaxy1999,
    coleHierarchicalGalaxyFormation2000, kangSemianalyticalModelGalaxy2005,
guoDwarfSpheroidalsCD2011, henriquesGalaxyFormationPlanck2015}, and
empirical models that aim to establish the connection between galaxies and
halos statistically and empirically
\citep[e.g.][]{jingSpatialCorrelationFunction1998,
    berlindHaloOccupationDistribution2003, yangConstrainingGalaxyFormation2003,
    valeLinkingHaloMass2004, kravtsovDarkSideHalo2004,
    zhengTheoreticalModelsHalo2005, mosterGalacticStarFormation2013,
luEmpiricalModelStar2014, behrooziUNIVERSEMACHINECorrelationGalaxy2019}. A
primary consensus on the galaxy-halo connection has been reached that the
stellar mass-halo mass relation is nonlinear, but is roughly described by a
double-power-law function \citep{yangConstrainingGalaxyFormation2003,
behrooziAVERAGESTARFORMATION2013, wechslerConnectionGalaxiesTheir2018}.
Such a relation has been used to indicate that the efficiency of converting
baryons into stars is suppressed in both the low-mass and high-mass ends,
respectively by supernova feedback
\citep[e.g.][]{dekelOriginDwarfGalaxies1986,
ceverinoRoleStellarFeedback2009} and active galactic nuclei (AGN) feedback
\citep[e.g.][]{crotonManyLivesActive2006,
fabianObservationalEvidenceAGN2012}, although the details remain uncertain.

Motivated by the success in establishing the primary connection between
galaxies and dark matter halos, attempts have been made to extend the
galaxy-halo connection using other (secondary) properties of galaxies and
halos. An important example is the attempt to relate the assembly histories of
halos to the star formation histories of galaxies using heuristic assumptions
that are tuned to match the observed abundance and spatial distribution of
galaxies as a function of their secondary properties, such as color and star
formation rate \citep[e.g.][]{hearinDarkSideGalaxy2013,
    hearinDarkSideGalaxy2014, watsonPredictingGalaxyStar2015,
    mosterEmergeEmpiricalModel2018,
    behrooziUNIVERSEMACHINECorrelationGalaxy2019,
wangRelatingGalaxiesDifferent2023, mengMeasuringGalaxyAbundance2020}.
Hydrodynamical simulations were also used to investigate the secondary
effects in the galaxy-halo connection. For example,
\citet{daviesQuenchingMorphologicalEvolution2020} found that central
galaxies living in early-formed halos tend to possess fewer amounts of gas
in the circum-galactic medium (CGM), harbor more massive black holes, have
longer cooling time for the CGM gas, and, consequently, are more quenched
in star formation. Such correlations indicate that early-formed halos tend
to host central galaxies that are more quiescent \citep[see
also][]{mattheeOriginScatterStar2019}. The causality between halo formation
and star formation status is further confirmed by
\citet{daviesQuenchingMorphologicalEvolution2021} using a controlled
simulation which follows the details of halo formation and star formation.
They found that early-formed halos grow their central black holes earlier
so that their feedback can quench the star formation activity earlier than
that in late-formed counterparts. In contrast,
\citet{cuiOriginGalaxyColour2021} found that early-formed halos actually
prefer to host star-forming central galaxies in their SIMBA simulation
\citep{daveMUFASAGalaxyFormation2016,
daveSIMBACosmologicalSimulations2019}. They argued that central galaxies in
early-formed halos are able to accumulate a large amount of cold gas
through cold-mode accretion at high-$z$
\citep{birnboimVirialShocksGalactic2003, keresHowGalaxiesGet2005,
vandevoortRatesModesGas2011, wrightRevealingPhysicalProperties2021} and
that the associated quasar-mode AGN feedback is inefficient to quench
galaxies but can extend the star formation time scale. On the other hand,
late-formed halos, which acquire their gas mainly through hot-mode
accretion, host central galaxies with smaller amounts of cold gas, and the
radio-mode AGN feedback associated with such accretion can quench these
galaxies. This scenario seems to be supported by observational results
where red central galaxies prefer to live in more massive halos (which tend
to form later) than their blue counterparts of the same stellar mass
\citep[e.g.][]{mandelbaumDensityProfilesGalaxy2006,
    mandelbaumStrongBimodalityHost2016, moreSatelliteKinematicsIII2011,
    wechslerConnectionGalaxiesTheir2018, postiPeakStarFormation2019,
    postiDynamicalEvidenceMorphologydependent2021,
zhangMassiveStarformingGalaxies2022}.

In order to study the impact of halo formation histories on central galaxies,
it is crucial to fix the halo mass and find an observational quantity that can
be used as a proxy of halo formation time.
\citet{wangInternalPropertiesEnvironments2011} found that the mass ratio
between the central subhalo and the host halo is strongly correlated with the
formation time of the host halo. Motivated by this and the expectation that the
stellar mass of a galaxy is correlated with the mass of its subhalo,
\citet{limObservationalProxyHalo2016} suggested that the central stellar mass
to halo mass ratio (hereafter central SMHMR) may be used as an observational
proxy of halo formation time \citep[see also][]{tojeiroGalaxyMassAssembly2017,
    bradshawPhysicalCorrelationsScatter2020,
zhangConnectionsGalaxyProperties2021a}. In this paper, we adopt the same proxy
and apply it to the observational data. We will show that central galaxies with
lower SMHMRs are more quenched. To make connections to halo formation time, we
calibrate the relation between central SMHMRs and halo formation time using
three different models, and they all agree with each other. We then use this
relation to quantify how the quenched fraction and the spiral fraction of
galaxies depend on the formation time of their host halos.

The paper is organized as follows. In
\S\,\ref{sec:testing_the_proxy_of_halo_formation_time} we describe the halo
formation time proxy and derive its relationship with halo formation time using
three different models, from the physical hydrodynamical simulation of
IllustrisTNG to the empirical subhalo halo abundance matching method. In
\S\,\ref{sec:applications_to_real_data}, we apply the proxy on the
observational data to study the dependence of central galaxy properties on the
halo formation time. Finally, our results are summarized in
\S\,\ref{sec:summary}. We converted all of the presented results to a
concordance $\Lambda$CDM cosmology with $H_0=100h~\rm km/s/Mpc$, $h=0.7$,
$\Omega_{\Lambda}=0.75$, and $\Omega_{m}=0.25$.

\section{Testing the proxy of halo formation time}%
\label{sec:testing_the_proxy_of_halo_formation_time}

\subsection{The proxy}%
\label{sub:the_proxy}

\begin{table*}
	\centering
	\caption{Notations used in this paper}
	\label{tab:notations}
	\begin{tabular}{cc} % four columns, alignment for each
		\toprule
        Notation & Description \\
         \midrule
        $V_{\rm peak}$ & Peak value of the maximum circular velocity of subhalos along
        the main branch.\\
         \midrule
        $t_{\rm form}$ & Lookback time when the FoF halo has first assembled half of its final
        mass along the main branch\\
         \midrule
        $\Delta t_{\rm form}$ & Residual of $t_{\rm form}$ with respect to the median $t_{\rm form}$
        in bins of halo mass\\
         \midrule
        central SMHMR & the stellar mass to halo mass ratio for central galaxies\\
         \midrule
        $M_{*, \rm sim}$ & Stellar mass of galaxies/subhalos in the hydrodynamical simulation\\
         \midrule
        $M_{*, \rm AM'}$ & \makecell[c]{Stellar mass assigned to each subhalo using the abundance matching
        method\\ according to the rank of $V_{\rm peak}$ based on the observational stellar mass function}\\
         \midrule
        $M_{h, \rm AM}$ & \makecell[c]{Halo mass assigned to each FoF halo using the abundance matching
        method\\ according to the rank of the sum of $M_{*, \rm sim}$ for all subhalos in this FoF halo}\\
         \midrule
        $M_{h, \rm AM'}$ & \makecell[c]{Halo mass assigned to each FoF halo using the abundance matching
        method\\ according to the rank of the sum of $M_{*, \rm AM'}$ for all subhalos in this FoF halo}\\
         \midrule
         \bottomrule
	\end{tabular}
\end{table*}

Previous studies suggest that the central SMHMR is a proxy of the halo
formation time, where halos that are formed earlier prefer to host more massive
central galaxies and vice versa \citep{wangInternalPropertiesEnvironments2011,
    limObservationalProxyHalo2016, tojeiroGalaxyMassAssembly2017,
    bradshawPhysicalCorrelationsScatter2020,
correaDependenceGalaxyStellartohalo2020}. Their relationship can be explained
by that early-formed halos have more time for their satellites to merge with
central galaxies.

A robust scaling relation between the central SMHMR and the halo formation time
needs to be built before we can use this proxy. To this end, we devised three
different models to quantify this scaling relation, as well as their scatters.
We start from a cosmological hydrodynamical simulation with main halos
identified where the halo mass is denoted as $M_{h,\rm sim}$. Each dark matter
halo contains no less than one subhalo and each subhalo contains one galaxy
with stellar mass assigned, which is denoted as $M_{*, \rm sim}$. Besides, each
subhalo also has a $V_{\rm peak}$, which is the peak value of the maximum
circular velocity along the evolution history of this subhalo. Then, we devised
three models with different assignments of stellar mass and halo mass to
subhalos and halos, i.e.
\begin{enumerate}

    \item \texttt{Model-A:} We use stellar mass and halo mass from the
        simulation, i.e. $M_{*, \rm sim}$ and $M_{h, \rm sim}$. The central
        galaxy is defined as the one with the largest $M_{*, \rm sim}$.

    \item \texttt{Model-B:} We use stellar mass for each subhalo in the
        simulation, i.e. $M_{*, \rm sim}$, and recalibrate the halo mass for
        each main halo with the abundance matching method according to the rank
        of the total stellar mass of member galaxies with $M_{*, \rm sim}\geq
        10^9M_{\odot}$, and this halo mass is denoted as $M_{h, \rm AM}$. The
        central galaxy is defined as the one with the largest $M_{*, \rm sim}$.

    \item \texttt{Model-C:} This model is fully empirical and independent of
        the specific implementation of baryonic physics. First, we assign
        stellar mass to each subhalo using the abundance matching method
        according to the rank of $V_{\rm peak}$, and this stellar mass is
        denoted as $M_{*, \rm AM'}$ \citep{reddickConnectionGalaxiesDark2013}.
        During the abundance matching procedure, the observational stellar mass
        function in \citet{moustakasPRIMUSCONSTRAINTSSTAR2013} is
        adopted\footnote{We note that the stellar mass function measurement in
            \citet{moustakasPRIMUSCONSTRAINTSSTAR2013} suffers from systematic
            effects in photometry. To maintain consistency in observational
            data, we adopt the re-calibrated results in
            \citet{behrooziUNIVERSEMACHINECorrelationGalaxy2019} where an
        attempt was made to deal with these issues.}. Then, we recalibrate
        the halo mass for each main halo with the abundance matching method
        according to the rank of the sum of $M_{*, \rm AM'}$ for all member
        galaxies with $M_{*, \rm AM'} \geq 10^9M_{\odot}$, and this halo
        mass is denoted as $M_{h, \rm AM'}$. The central galaxy is defined
        as the one with the largest $M_{*, \rm AM'}$.

\end{enumerate}
These notations are summarized in Table~\ref{tab:notations}. We note that
\texttt{Model-A} totally relies on the hydrodynamical simulation, so the
stellar mass and halo mass are both physical, but depend on the specific
implementation of the simulation used. On the contrary, \texttt{Model-C} only
relies on the heuristic assumption of abundance matching. And \texttt{Model-B}
is in between. Using these three models, from the physical hydrodynamical
simulation to the empirical statistical model, we will demonstrate that they
all agree on the scaling relation between the central SMHMR and the formation
time of their host halos, showing the robustness of this relation.

\subsection{Numerical simulations used for the test}%
\label{sub:numerical_simulations_used_for_the_test}

Here we use the state-of-the-art simulation of IllustrisTNG to test the scaling
relation between central SMHMRs and halo formation time. The IllustrisTNG (The
Next Generation) \citep[hereafter
TNG,][]{weinbergerSimulatingGalaxyFormation2017,
    pillepichSimulatingGalaxyFormation2018,
    springelFirstResultsIllustrisTNG2018, naimanFirstResultsIllustrisTNG2018,
    marinacciFirstResultsIllustrisTNG2018,
    pillepichFirstResultsIllustrisTNG2018, nelsonFirstResultsIllustrisTNG2018,
nelsonIllustrisTNGSimulationsPublic2019} is a suite of
gravo-magnetohydrodynamical cosmological simulations run with the
moving-mesh code AREPO. TNG simulates the formation and evolution of
galaxies from $z=127$ to $z=0$ based on a cosmology consistent with results
in \citet{planckcollaborationPlanck2015Results2016}, where
$\Omega_{\Lambda, 0}=0.6911$, $\Omega_{b, 0}=0.3089$, $\sigma_8=0.8159$,
$n_s=0.9667$ and $h=0.6774$. Here we use TNG300-1 for better statistics,
which has $2\times 2500^3$ resolution elements in a box with the side
length of $205h^{-1}{\rm cMpc}$. The target baryon mass resolution is
$1.1\times 10^7 M_{\odot}$, and the dark matter particle mass is $5.9\times
10^7 M_{\odot}$.

In the TNG simulation, dark matter halos were identified with the
friends-of-friends (FoF) algorithm using dark matter particles\footnote{These
    halos are also called main halos or FoF halos, and we will use them
interchangeably.}. Substructures were identified with the SUBFIND algorithm
\citep{springelPopulatingClusterGalaxies2001} using all types of particles,
where the baryonic components identified are defined as galaxies, with the dark
matter components as subhalos. The subhalo located at the minimum of the
gravitational potential is defined as the central one, while others are
satellites. Subhalo merger trees are built using the SUBLINK algorithm
\citep{rodriguez-gomezMergerRateGalaxies2015}. Once all the trees are built up,
we can define the main progenitor of any subhalo in the preceding snapshot as
the one with {\it the most massive progenitor history}
\citep[see][]{deluciaHierarchicalFormationBrightest2007}. Finally, we can
obtain a branch consisting of the main progenitors of the halo in question by
recursively identifying the main progenitor, and this branch is called the main
branch.

The halo mass is defined as the total mass within a sphere around the halo
center where the mean overdensity is 200 times the critical density of the
universe. The peak halo mass of a subhalo is defined as the maximum halo mass
that the subhalo has ever achieved along the main branch when it is identified
as a central subhalo. For each subhalo, we can also calculate its maximum
circular velocity, and denote it as $V_{\rm max}$. Then, we trace the main
branch of this subhalo to find the peak value for all $V_{\rm max}$ on this
branch and denote it as $V_{\rm peak}$.

The assembly history of each dark matter halo corresponds to a subhalo merger
tree, where small halos are formed early, then they merge with each other and
end up with one descendant halo at $z=0$. Despite the complexity of this merger
tree, we can still capture the main feature with the characteristic formation
time of halos \citep{chenRelatingStructureDark2020}. There are many different
definitions of halo formation time \citep[e.g.][]{gaoAgeDependenceHalo2005,
    wechslerDependenceHaloClustering2006, liHaloFormationTimes2008,
nadlerSymphonyCosmologicalZoomin2023}, and here we choose to use the simplest
one, which is defined as the highest redshift when the main progenitor halo has
assembled half of its descendant halo mass at $z=0$. The corresponding lookback
time is denoted as $t_{\rm form}$.

\subsection{Test results}%
\label{sub:test_results}

\begin{figure*}
    \centering
    \includegraphics[width=0.9\linewidth]{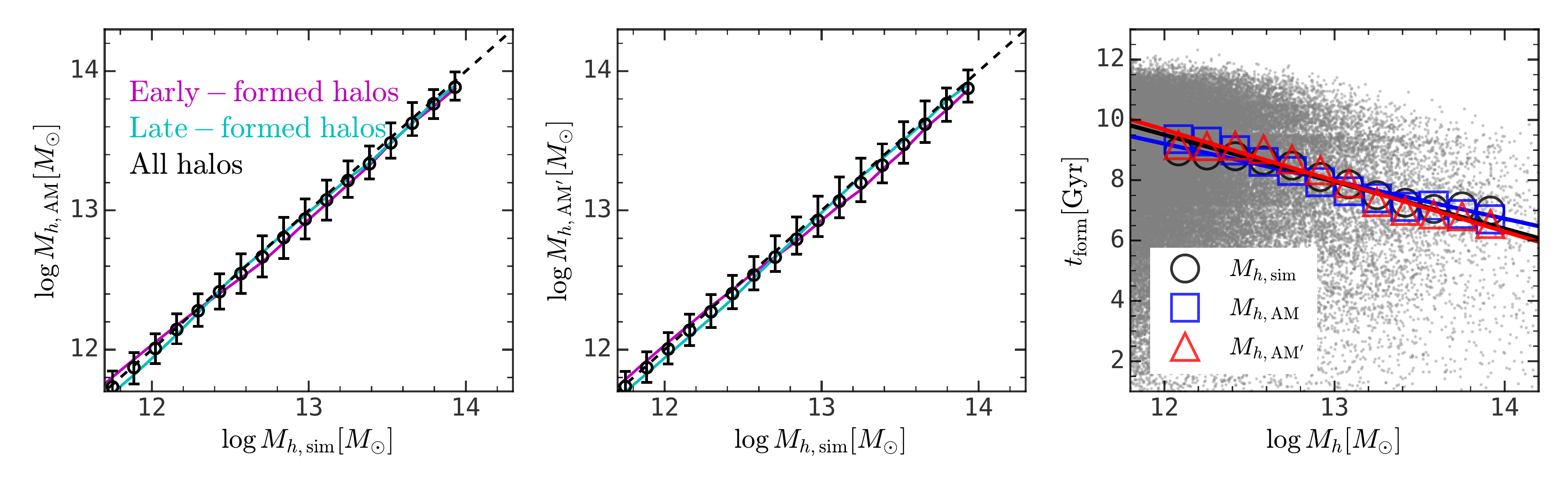}
    \caption{
        {\bf Left panel:} Comparison of halo mass between the TNG300
        simulation, $M_{h, \rm sim}$, and the abundance matching result, $M_{h,
        \rm AM}$, according to the rank of the sum of $M_{*, \rm sim}$ in each
        group. {\bf Middle panel:} Comparison of halo mass between the TNG300
        simulation, $M_{h, \rm sim}$, and the abundance matching result, $M_{h,
        \rm AM'}$, according to the rank of the sum of $M_{*, \rm AM'}$ in each
        group. In both panels, circles with error bars show the median value in
        each mass bin with errors estimated using the bootstrap method. The
        dashed line is the one-to-one reference line. Magenta and cyan lines
        are for early-formed and late-formed halos, respectively. {\bf Right
        panel:} Relation between halo mass and halo formation time. The
        background points are the TNG halos. The black/blue/red symbols are for
        $M_{h, \rm sim}$/$M_{h, \rm AM}$/$M_{h, \rm AM'}$, respectively. In
        this paper, we only use galaxies with stellar mass above
        $10^9M_{\odot}$ in halos with halo mass above $10^{12}M_{\odot}$ for
        each definition of stellar mass and halo mass. 
    }%
    \label{fig:figures/sham_performance_tng300}
\end{figure*}

\begin{figure*}
    \centering
    \includegraphics[width=0.9\linewidth]{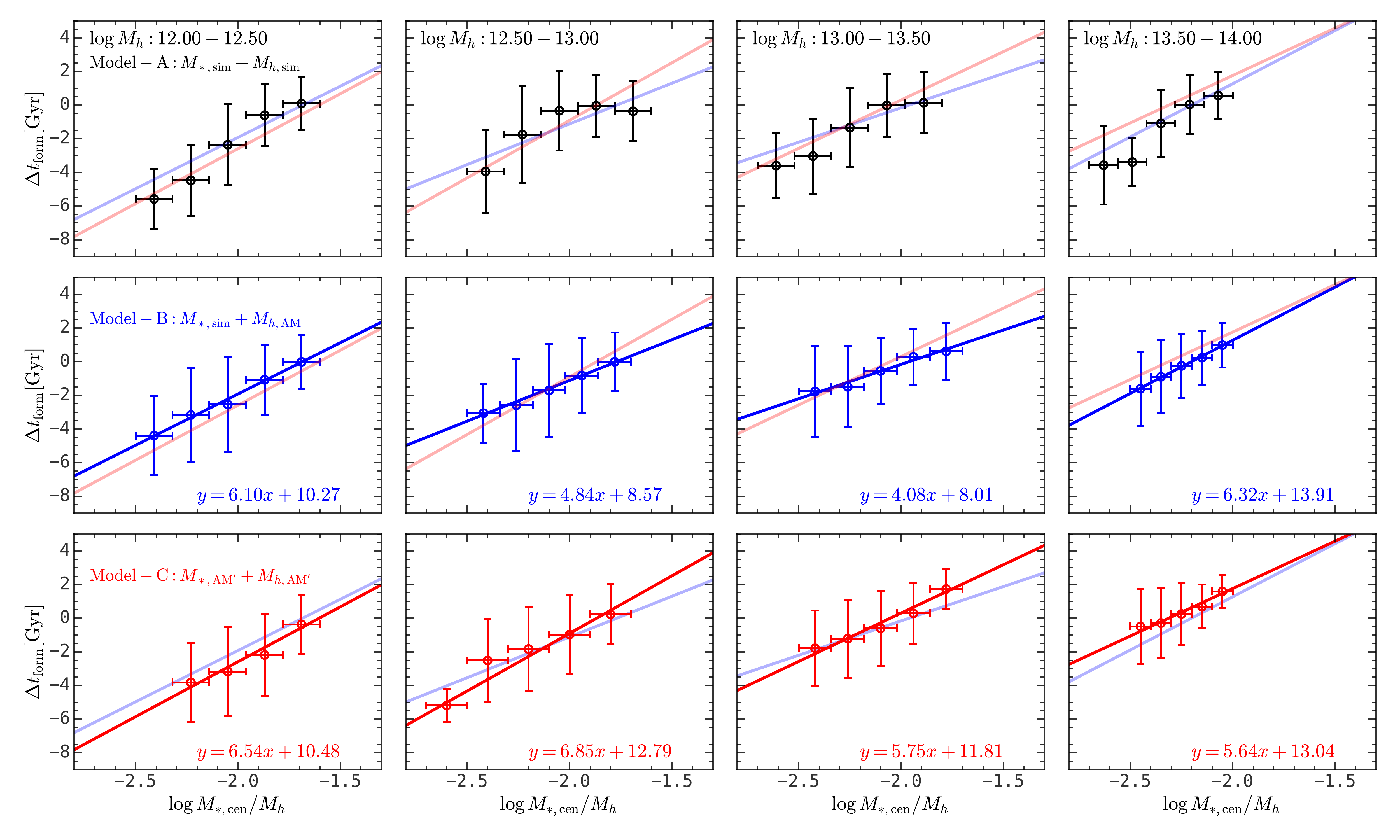}
    \caption{
        Relation between halo formation time, $\Delta t_{\rm form}$, and the
        central SMHMR, $\log(M_{*,\rm cen}/M_h)$. Symbols are the mean $\Delta
        t_{\rm form}$ in each central SMHMR bin with error bars showing the
        standard deviation. The blue/red solid lines are the linear fitting
        results for data points on the middle/bottom panels, respectively. {\bf
        Top panels:} Both stellar mass and halo mass are from the TNG300
        simulation, i.e. \texttt{Model-A}. {\bf Middle panels:} The stellar
        mass is from the TNG300 simulation, and the halo mass is recalibrated
        by rank-matching the total stellar mass in each group, i.e.
        \texttt{Model-B}. {\bf Bottom panels:} The stellar mass is recalibrated
        with the abundance matching method according to the rank of $V_{\rm
        peak}$ for each subhalo, and the halo mass is recalibrated by
        rank-matching the total stellar mass in each group, i.e.
        \texttt{Model-C}. This figure illustrates that halos with larger
        central SMHMRs are formed early.
    }%
    \label{fig:figures/smhmr_dtform_relation_tng300_am}
\end{figure*}

\begin{figure*}
    \centering
    \includegraphics[width=0.9\linewidth]{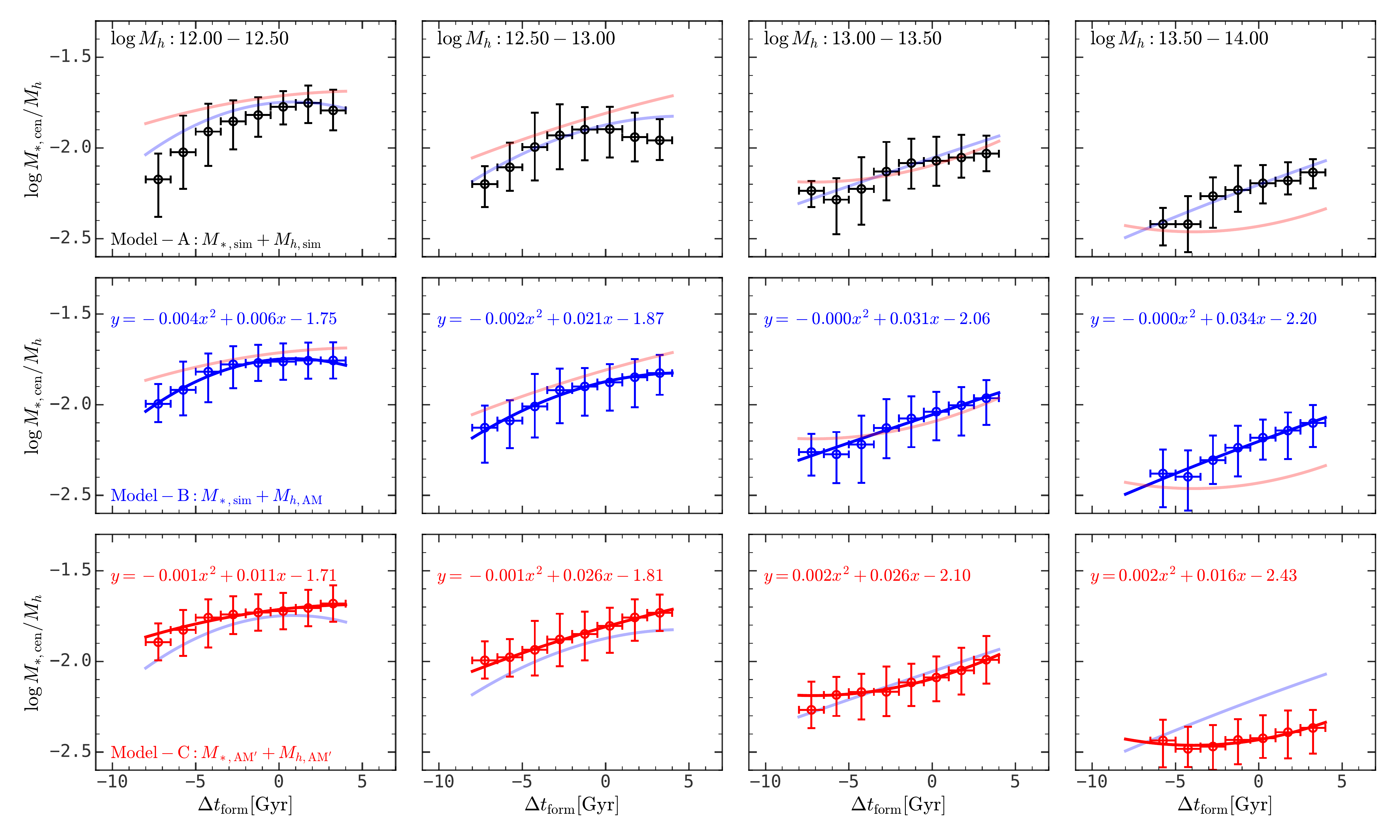}
    \caption{
        Relation between central SMHMR, $\log(M_{*, \rm cen}/M_h)$, and halo
        formation time, $\Delta t_{\rm form}$. Symbols are the median central
        SMHMR in each $\Delta t_{\rm form}$ bin with error bars showing the
        16\%-84\% quantiles. The blue/red solid lines are the fitting results
        for data points on the middle/bottom panels, respectively. The stellar
        mass and halo mass for panels on different rows are the same as
        Fig.~\ref{fig:figures/smhmr_dtform_relation_tng300_am}. This figure
        illustrates that early-formed halos tend to have larger central SMHMRs,
        i.e. more massive central galaxies.
    }%
    \label{fig:figures/dtform_smhmr_relation_tng300_am}
\end{figure*}

\subsubsection{Comparison of three models}%
\label{ssub:comparison_of_three_models}

We first examine the difference in halo masses for these three models. The left
panel in Fig.~\ref{fig:figures/sham_performance_tng300} shows the comparisons
between $M_{h, \rm sim}$ and $M_{h, \rm AM}$. Here the circles with error bars
show the median and $16\%-84\%$ quantiles, and the dashed line is the
one-to-one reference line. Similarly, the middle panel shows the comparison
between $M_{h, \rm sim}$ and $M_{h, \rm AM'}$. All these results show that the
halo masses assigned according to the empirical abundance matching method are
very close to the values in the TNG300 simulation.

Then, we examine the relationship between three halo masses and halo formation
time, $t_{\rm form}$, which is shown on the right panel in
Fig.~\ref{fig:figures/sham_performance_tng300}. The symbols are the median
$t_{\rm form}$ in each halo mass bin. The solid lines are the corresponding
linear fitting line, while the black/blue/red colors are for three models
accordingly. Here one can see that all three models produce a nearly identical
negative correlation between halo mass and halo formation time, where massive
halos are formed late, which is expected in the hierarchical structure
formation scenario. We note that this correlation can introduce bias into our
study when a finite halo mass bin is used. Hence, we use the residual with
respect to the median $t_{\rm form}$ as a function of halo mass, i.e.
\begin{equation}
    \Delta t_{\rm form} = t_{\rm form} - \texttt{Median}(t_{\rm form}(M_h)).
    \label{eq:dtform}
\end{equation}

Finally, to see if the abundance matching method introduces any dependence on
the halo formation time, we compare the halo mass in three models for
early-formed and late-formed halos separately, where halos with $\Delta t_{\rm
form} > 0$ in each $M_{h,\rm sim}$ bin are defined as early-formed and others
are late-formed. The results are shown in the first three panels of
Fig.~\ref{fig:figures/sham_performance_tng300} with magenta and cyan solid
lines, respectively. Here one can see that our abundance matching method does
not introduce any obvious bias.

\subsubsection{The central SMHMR-halo formation time relation}%
\label{ssub:the_central_smhmr_halo_formation_time_relation}

Fig.~\ref{fig:figures/smhmr_dtform_relation_tng300_am} shows the mean $\Delta
t_{\rm form}$ as a function of central SMHMRs, as well as the standard
deviation. The top/middle/bottom panels show the results in
\texttt{Model-A}/\texttt{Model-B}/\texttt{Model-C}. Here one can see that all
three models predict very similar scaling relations that halos with higher
central SMHMR are formed earlier. These results indicate that the central
SMHMR-halo formation time relation does not depend on the specific
implementation of baryonic physics in hydrodynamical simulations. Instead, this
correlation can be explained by the hierarchical structure formation scenario,
where early-formed halos have more time for their central galaxies to merge
with their satellites.  We also fit the scaling relations between central
SMHMRs and $\Delta t_{\rm form}$ with linear functions for \texttt{Model-B} and
\texttt{Model-C}, and present them with blue and red lines, respectively.

Fig.~\ref{fig:figures/dtform_smhmr_relation_tng300_am} shows the median central
SMHMR as a function of $\Delta t_{\rm form}$ in four halo mass bins for three
models. The scaling relations in \texttt{Model-B} and \texttt{Model-C} are
fitted with quadratic polynomials, which are shown in blue and red solid
curves. Here one can see that all three models predict that early-formed halos
have higher central SMHMRs and the scaling relations are similar, except for
massive halos in \texttt{Model-C}. The discrepancies between \texttt{Model-C}
and \texttt{Model-B}/\texttt{Model-A} are attributed to the difference in their
stellar mass functions, where TNG300 underestimated the amplitude of stellar
mass functions due to its relatively low resolution
\citep{pillepichFirstResultsIllustrisTNG2018}. We also checked the relation
using the EAGLE simulation and presented the result in
appendix~\ref{sec:eagle}, which agrees well with our results here.

It is noteworthy that our results on the positive correlation between halo
formation time and SMHMRs is consistent with previous studies on galaxy
assembly bias \citep[e.g.][]{zentnerGalaxyAssemblyBias2014,
    hearinIntroducingDecoratedHODs2016, maoAssemblyBiasExploring2018,
wangHowOptimallyConstrain2019, zuDoesConcentrationDrive2021}, which is the
dependence of the halo occupation of galaxies on secondary properties of dark
matter halos manifested in their spatial distributions.
\citet{zentnerConstraintsAssemblyBias2019} found that high-concentration halos
prefer to host more luminous central galaxies compared with their
low-concentration counterparts of the same halo mass through a comprehensive
Bayesian analysis; a similar conclusion is obtained in
\citet{wangEvidenceGalaxyAssembly2022} by combining several spatial summary
statistics. In this paper, we use the group-based statistic, i.e. SMHMR, and
obtain consistent results based on three different models, supporting previous
findings.

\section{Applications to real data}%
\label{sec:applications_to_real_data}

In the previous section, we established the relationship between central SMHMRs
and halo formation time using three different models, from the physical
hydrodynamical simulation to the empirical statistical model. All these models
produce similar results that halos with higher central SMHMRs are formed
earlier, and vice versa. These results illustrate that the relation between
halo formation time and the central SMHMRs does not rely on the specific
implementation of galaxy formation models nor specific subgrid recipes (see
Appendix~\ref{sec:eagle}). With these scaling relations, we can use the central
SMHMR as a proxy of halo formation time and study its relation with central
galaxy properties in our real Universe.

\subsection{Observational data}%
\label{sub:observational_data}

\begin{figure*}
    \centering
    \includegraphics[width=1\linewidth]{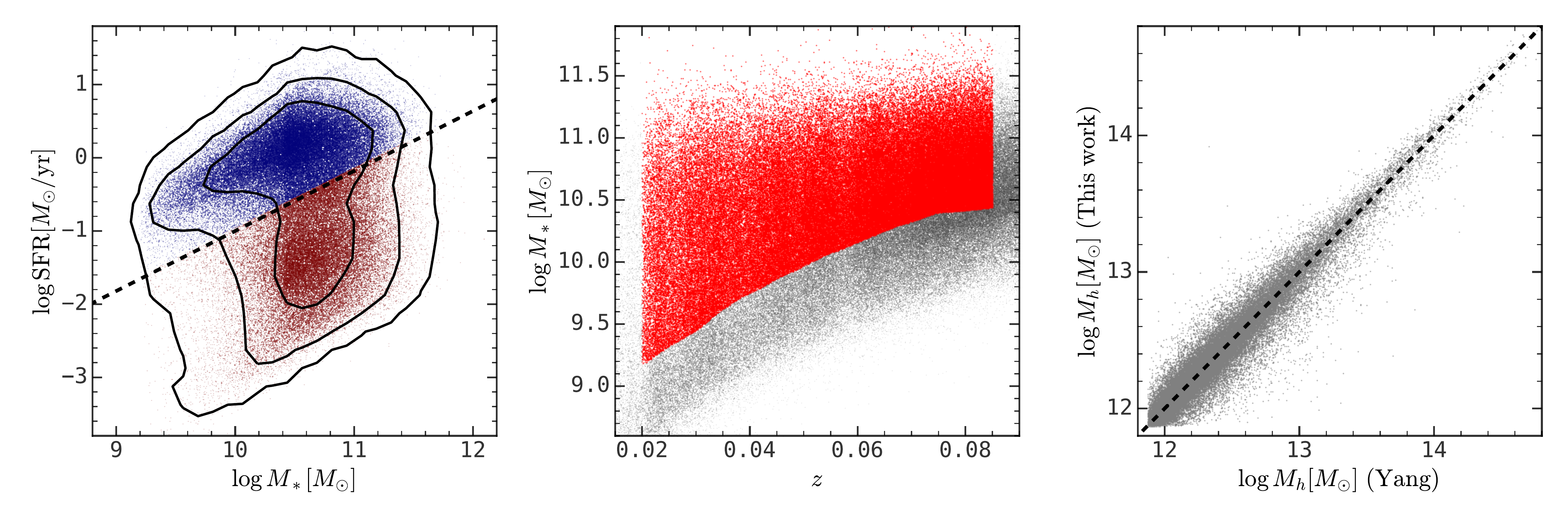}
    \caption{
        {\bf Left panel:} The distribution of galaxies on the stellar mass and
        SFR plane with black solid lines enclose 68\%, 95\%, and 99.7\% of the
        whole sample. The dashed line is the calibrated separation line for the
        star-forming and quiescent galaxies. The red points are quiescent
        galaxies and the blue ones are star-forming galaxies. {\bf Middle
        panel:} The distribution of galaxies on the redshift and stellar mass
        plane, where red points are the selected mass-limited galaxies and gray
        points are the remaining ones. {\bf Right panel:} The comparison
        between the halo mass calibrated in \citet{yangGalaxyGroupsSDSS2007}
        and our recalibrated halo mass using the stellar mass from the GSWLC
        catalog (see equation~(\ref{eq:am})), and the black dashed line is the
        one-to-one reference line.
    }%
    \label{fig:figures/sdss_vlim_sample}
\end{figure*}

Here we use the GALEX-SDSS-WISE Legacy Catalog (GSWLC)
\citep{salimGALEXSDSSWISE2016, salimDustAttenuationCurves2018}, which is
constructed from the Sloan Digital Sky Survey (SDSS) main galaxy sample (MGS)
\citep{yorkSloanDigitalSky2000, blantonNewYorkUniversity2005,
abazajianSeventhDataRelease2009}. The SDSS MGS is a magnitude-limited
spectroscopic survey with $r < 17$ and covers $\sim 8000~\rm deg^2$. The
stellar mass and the star formation rate (SFR) are estimated from the
UV-optical-IR bands photometry using the CIGALE code
\citep{nollAnalysisGalaxySpectral2009, boquienCIGALEPythonCode2019} with the
stellar library of \citet{bruzualStellarPopulationSynthesis2003} and the
initial mass function of \citet{chabrierGalacticStellarSubstellar2003}. The
left panel of Fig.~\ref{fig:figures/sdss_vlim_sample} shows the distribution of
galaxies on the $M_*-\rm SFR$ plane. We further separate galaxies into the
star-forming and quiescent populations according to that whether the SFR is
more than 1 dex below the star-forming main sequence
\citep[see][]{2023arXiv230406886W, wooDependenceGalaxyQuenching2013}. The
separation line is
\begin{equation}
    \log \left(\frac{\rm SFR}{M_{\odot}/{\rm yr}}\right) = 0.82\times \log
    \left(\frac{M_*}{M_{\odot}}\right) - 9.17,
\end{equation}
and shown as the black dashed line on the left panel of
Fig.~\ref{fig:figures/sdss_vlim_sample}.

Based on the GSWLC catalog, we constructed a $M_*$-limited sample using the
method in \citet{2023arXiv230406886W} \citep[see
also][]{pozzettiZCOSMOS10kbrightSpectroscopic2010}. They are shown in red
points on the right panel of Fig.~\ref{fig:figures/sdss_vlim_sample} where the
black dashed line shows the minimal stellar mass we can probe at each redshift.
We also weigh each galaxy with $1/V_{\rm max}$ where $V_{\rm max}$ is the
volume between $z_{\rm min}=0.02$ and $z_{\rm max}$, which is inferred from the
mass-limited envelope calculated above according to the stellar mass of each
galaxy.

We use the morphology classification from the Galaxy Zoo
project\footnote{https://data.galaxyzoo.org/} \citep{lintottGalaxyZooData2011}.
Through Galaxy Zoo, galaxies are classified into different categories of
morphology according to their visual image by citizen scientists, and each
galaxy has more than 20 votes. Then a flag, i.e. \texttt{spiral},
\texttt{elliptical}, or \texttt{uncertain}, is assigned to each galaxy
according to the note frequency corrected with a debiasing process
\citep{bamfordGalaxyZooDependence2009}. Here we define spiral galaxies as those
with the flag of \texttt{SPIRAL} set to unity.

We use the group catalog of \citet{yangGalaxyGroupsSDSS2007} to select central
galaxies as the most massive one in each galaxy group. The halo mass is
calibrated by abundance-matching the total stellar mass with theoretical halo
mass function, where the stellar mass estimation is based on
\citet{bellOpticalInfraredProperties2003}. For the consistency of the stellar
mass estimation, we recalibrate the halo mass for each group using the
abundance matching method according to the rank of the sum of stellar mass
estimated in \citet{salimGALEXSDSSWISE2016}, which is
\begin{equation}
    n_h(>M_h) = n_*(>M_{*, \rm tot}) \label{eq:am}
\end{equation}
where the left hand is the cumulative halo mass function and the right hand is
the cumulative total stellar mass function. The right panel of
Fig.~\ref{fig:figures/sdss_vlim_sample} shows the comparison between the halo
mass in \citet{yangGalaxyGroupsSDSS2007} and our recalibrated ones. In
appendix~\ref{sec:dependence_tform_on_central}, we demonstrated the importance
of the halo mass recalibration to avoid bias, especially for low-mass groups.

\subsection{The dependence of central galaxy properties on halo formation time}%
\label{sub:dependence_smhmr_tform}

\begin{figure*}
    \centering
    \includegraphics[width=0.9\linewidth]{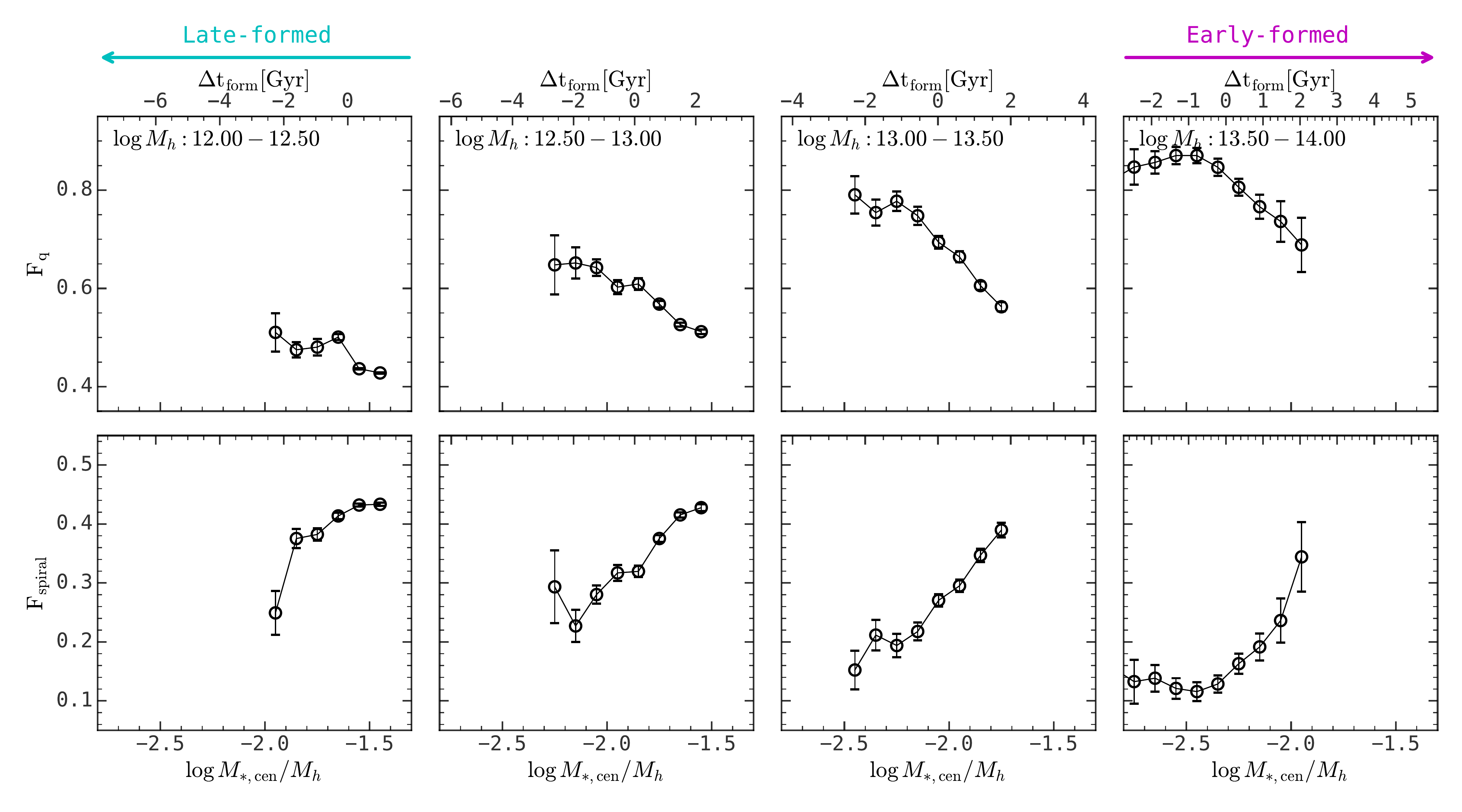}
    \caption{
        The quiescent fraction ({\bf top panels}) and the spiral fraction ({\bf
        bottom panels}) of central galaxies as a function of central SMHMRs in
        different halo mass bins. Error bars are estimated with the bootstrap
        method. The top x-axis shows the halo formation time converted from
        central SMHMRs using the scaling relation calibrated in
        \texttt{Model-C} (see bottom panels in
        Fig.~\ref{fig:figures/smhmr_dtform_relation_tng300_am}). Halos with
        larger $\Delta t_{\rm form}$ are formed earlier, as indicated by the
        arrows on the figure. This figure illustrates that central galaxies in
        halos with high SMHMRs, which are formed early, have low quiescent
        fractions and higher spiral fractions.
    }%
    \label{fig:figures/ssfr_wrt_csmr_hm_bins_sdss}
\end{figure*}

\begin{figure*}
    \centering
    \includegraphics[width=0.9\linewidth]{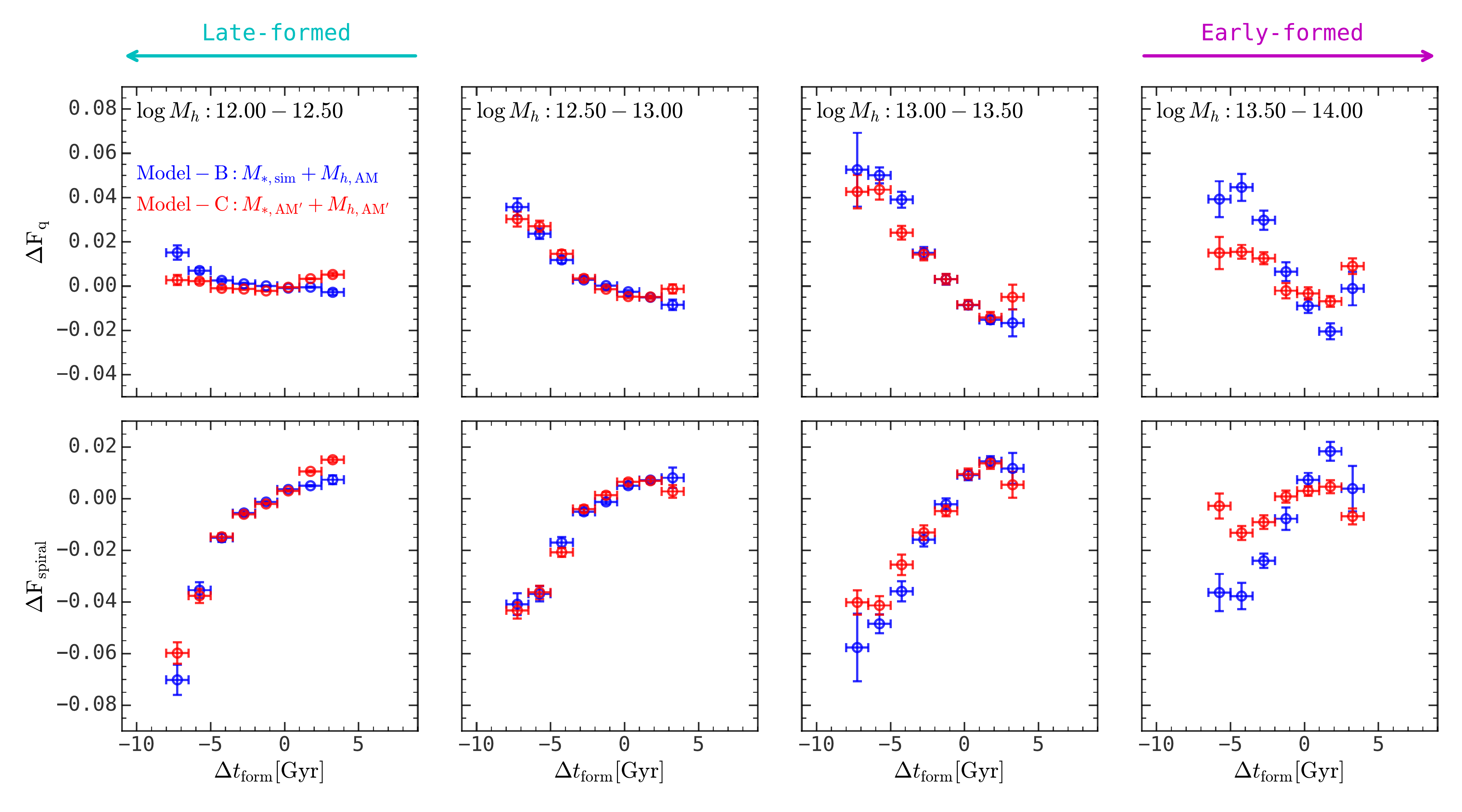}
    \caption{
        Dependence of the quiescent fraction and the spiral fraction on $\Delta
        t_{\rm form}$ derived from equation~(\ref{eq:stats}). Symbols show the
        mean deviation from the mean fraction in each halo mass bin with error
        bars estimated using the bootstrap method. This figure illustrates that
        central galaxies in late-formed halos have higher quiescent fractions
        and lower spiral fractions.
    }%
    \label{fig:figures/ssfr_wrt_dtform_hm_bins_sdss_2}
\end{figure*}

Fig.~\ref{fig:figures/ssfr_wrt_csmr_hm_bins_sdss} shows the quiescent fraction
and the spiral fraction of central galaxies as a function of central SMHMRs in
four halo mass bins. As one can see, for the whole halo mass range, central
galaxies with low SMHMRs are more quenched and have lower spiral fractions than
those with high SMHMRs. The difference in the quiescent fraction ranges from
$\sim 10\%$ to $\gtrsim 20\%$, while the difference in the spiral fraction is
about $\sim 20\%$ across the whole halo mass range in question. It is
noteworthy that we are using a broad halo mass bin width for better statistics.
We also checked the results using finer halo mass bins in
appendix~\ref{sec:finer_halo_mass_bin}, and the trend is consistent with the
results shown here.

On the top x-axis of Fig.~\ref{fig:figures/ssfr_wrt_csmr_hm_bins_sdss}, we also
show $\Delta t_{\rm form}$, which is converted from the central SMHMR using the
scaling relation calibrated in \texttt{Model-C} (see bottom panels in
Fig.~\ref{fig:figures/smhmr_dtform_relation_tng300_am}). One can see that halos
with low central SMHMRs, which are formed late, are not only more quiescent,
but also have a lower fraction of spiral morphology. The range of the halo
formation time we can probe is $\sim 4$ Gyr in each halo mass bin. We note
that the results in Fig.~\ref{fig:figures/ssfr_wrt_csmr_hm_bins_sdss} cannot be
simply interpreted as the dependence of the quiescent fraction and the spiral
fraction on the halo formation time, because the scaling relation between
$\Delta t_{\rm form}$ and the central SMHMR has a non-negligible scatter.

To derive the dependence of central galaxy properties on the formation time of
their host halos more strictly, we must consider the scatter in the scaling
relation of central SMHMRs and halo formation time. To this end, we can express
the distribution of central galaxy properties, $F$, as a function of halo
formation time, $\Delta t_{\rm form}$, as
\begin{align}
    P(F\mid \Delta t_{\rm form}) &= \int dR P(F, R\mid \Delta t_{\rm
    form})\nonumber\\ &= \int dR P(F \mid R, \Delta t_{\rm form})P(R\mid \Delta
    t_{\rm form})\nonumber\\
                      &\approx \int dR P(F\mid R)P(R\mid \Delta t_{\rm form}),
                      \label{eq:stats}
\end{align}
where $F$ represents the quiescent fraction or the spiral fraction, $R$ is the
central SMHMR, and $\Delta t_{\rm form}$ is the halo formation time defined in
equation~(\ref{eq:dtform}). We note that all of the probability distributions
above are conditioned on halo mass, and we omit it for clarity. The final step
of the above derivation is based on the assumption that $P(F\mid R, \Delta
t_{\rm form}) \approx P(F\mid R)$, which assumes that the impact of the halo
formation time on the star formation activities and morphologies of central
galaxies is only through the central SMHMR.

Based on the above derivation, we can calculate the quiescent fraction and the
spiral fraction of central galaxies as a function of the formation time of
their host halos. We start with all of the FoF halos in the TNG300 simulation,
where each halo has a halo mass and a central stellar mass assigned in
\texttt{Model-B} (or \texttt{Model-C}), as well as a halo formation time,
$\Delta t_{\rm form}$, calculated with equation~(\ref{eq:dtform}). For each
halo, we can calculate its central SMHMR, $R$, which is used to read a value of
mean quiescent/spiral fraction $F(R)$ and the associated scatter $\sigma_F(R)$
from Fig.~\ref{fig:figures/ssfr_wrt_csmr_hm_bins_sdss} in the corresponding
halo mass bin. Then we generate a quiescent/spiral fraction for the halo in
question from a normal distribution with a mean of $F(R)$ and a scatter of
$\sigma_F(R)$. Finally, we can calculate the mean quiescent/spiral fraction for
halos with given halo masses and $\Delta t_{\rm form}$. The result is shown in
Fig.~\ref{fig:figures/ssfr_wrt_dtform_hm_bins_sdss_2}, which shows the residual
with respect to the mean quiescent/spiral fraction in each halo mass bin. The
blue (red) symbols are results with \texttt{Model-B} (\texttt{Model-C}), where
both models produce very similar results. Here one can see that central
galaxies in late-formed halos are more quiescent and have a lower fraction of
spiral morphology than their early-formed counterparts. The difference in the
quiescent fraction is about $2\%-8\%$ for halos with $M_h >
10^{12.5}M_{\odot}$, and the difference in the spiral fraction is about $2\%-
8\%$ across the whole halo range in question.

The relation between halo formation time and central galaxy properties can also
manifest itself as the difference in the formation time for halos with
star-forming/spiral and quiescent/non-spiral central galaxies. We present the
result in appendix~\ref{sec:dependence_tform_on_central}, which shows that
halos with star-forming/spiral central galaxies are formed earlier than their
quiescent/non-spiral counterparts by $\lesssim 0.4$ Gyr.

\subsection{Impact of the group finding algorithm}%
\label{sub:impact_of_the_group_finding_algorithm}

Since dark matter halos are difficult to be detected observationally, we rely
on the group finding algorithm to construct the group catalog
\citep[e.g.][]{ekeGalaxyGroups2dFGRS2004, yangHalobasedGalaxyGroup2005,
    berlindPercolationGalaxyGroups2006, knobelZCOSMOS20kGROUP2012,
wangIdentifyingGalaxyGroups2020}. The performance of the halo-based group
finder is tested in many previous studies
\citep[e.g.][]{yangHalobasedGalaxyGroup2005, yangGalaxyGroupsSDSS2007,
    luGALAXYGROUPS2MASS2016, limGalaxyGroupsLowredshift2017,
yangExtendedHalobasedGroup2021}, which shows that it can recover the input
galaxy groups with high completeness and purity, and calibrate the halo mass of
galaxy groups to high accuracy, based on the mock test. Nevertheless, this group
finder is poorly tested on the performance in recovering color/SFR-related
statistics \citep{campbellAssessingColourdependentOccupation2015}, like the
quiescent fraction, which might jeopardize our conclusions in this paper. Here
we want to examine if this contamination can bias our results. To this end, we
applied the group finding algorithm to a mock galaxy catalog and see whether
the relation between central SMHMRs and quiescent fractions of central galaxies
is biased.

The mock survey is constructed based on the N-body simulation of ELUCID
\citep{wangELUCIDEXPLORINGLOCAL2014, wangELUCIDEXPLORINGLOCAL2016}, which is a
constrained simulation with a box volume of $(500h^{-1}{\rm Mpc})^3$ and
$3072^3$ dark matter particles. Dark matter halos are identified with the FoF
method and subhalos are identified with the SUBFIND algorithm
\citep{springelPopulatingClusterGalaxies2001}. The stellar mass and star
formation rate are assigned to each subhalo according to the empirical model of
MAHGIC, which is trained on the IllustrisTNG simulation \citep[see][for more
details]{chenHowEmpiricallyModel2021, chenMAHGICModelAdapter2021}. Here we only
use galaxies with $M_*\geq 10^9M_{\odot}$. The most massive galaxy in each FoF
halo is defined as the central one, while others are satellites. The
redshift-space position for each galaxy is generated using the method in
\citet{chenELUCIDVICosmic2019}.

\begin{figure}
    \centering
    \includegraphics[width=0.9\linewidth]{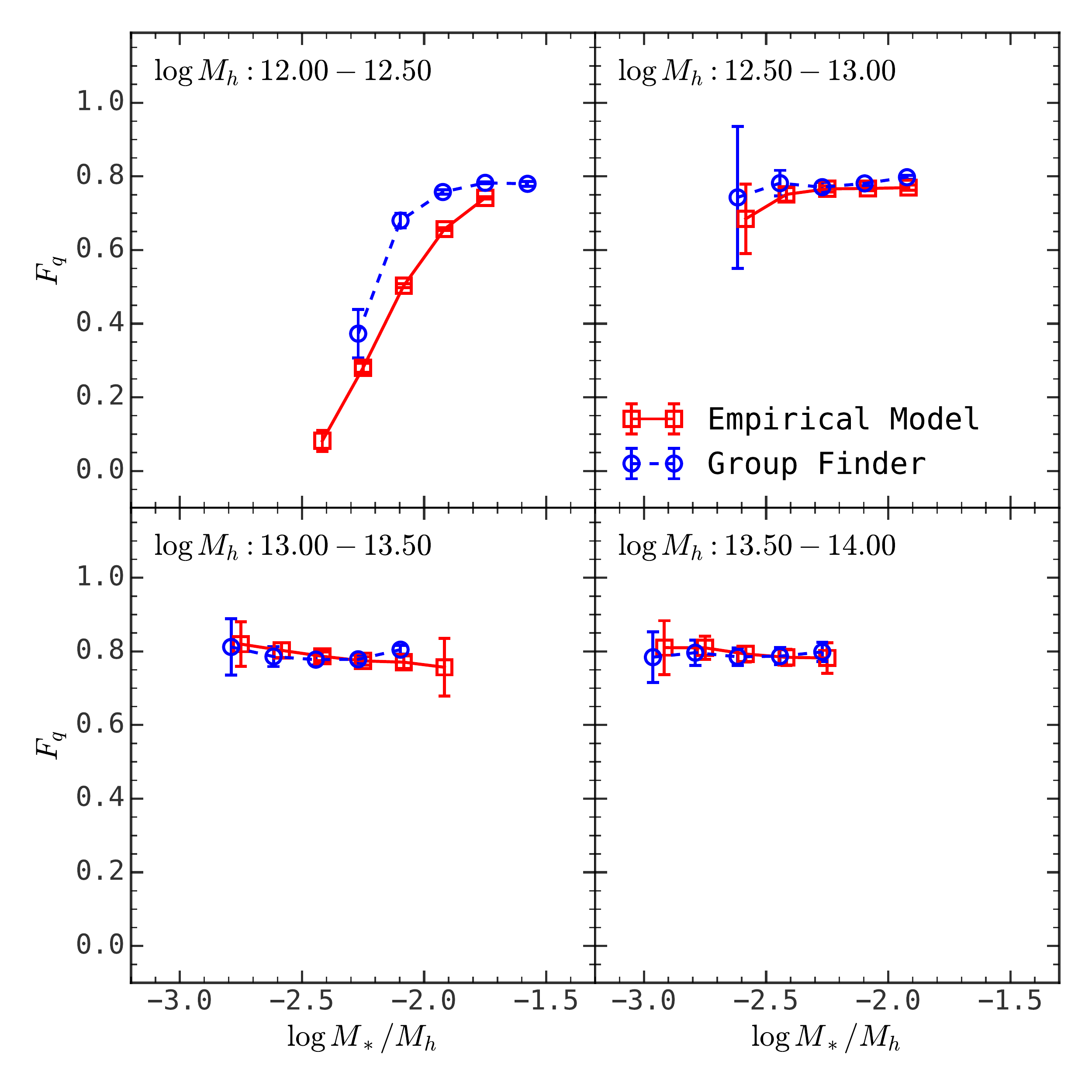}
    \caption{
        The quiescent fraction as a function of central SMHMRs in the empirical
        model of \texttt{MAHGIC} applied on the ELUCID simulation. The red
        color shows the results of the input catalog. The blue color shows the
        result of the halo-based group
        finder. Poisson errors are shown. This figure illustrates that the
        group finding algorithm does not bias the dependence of the quiescent
        fraction of central galaxies on the central SMHMRs.
    }%
    \label{fig:figures/mock_group_finder_result}
\end{figure}

For each FoF halo in the mock volume, we re-assign its halo mass using the
abundance matching method according to the rank of the total stellar mass in
this halo (see equation~(\ref{eq:am})). Then we calculate the fraction of
quiescent galaxies, which is defined as those with ${\rm SFR}/M_* <
10^{-11}{\rm yr}^{-1}$, as a function of central SMHMR in four halo mass bins.
The results are shown in red squares in
Fig.~\ref{fig:figures/mock_group_finder_result}.

Then, we applied the halo-based group-finding algorithm in
\citet{yangExtendedHalobasedGroup2021} to these mock galaxies and constructed a
galaxy group catalog \citep[see also][]{luGALAXYGROUPS2MASS2016,
    limGalaxyGroupsLowredshift2017, yangExtendedHalobasedGroup2021,
liGroupsProtoclusterCandidates2022}. For each identified galaxy group, the halo
mass is assigned with the abundance matching method according to the rank of
the total stellar mass in each group, and the central galaxy is defined as the
one with the largest stellar mass. Finally, we can plot the quiescent fraction
as a function of central SMHMRs in four halo mass bins based on the identified
galaxy group catalog, and the result is shown with blue circles in
Fig.~\ref{fig:figures/mock_group_finder_result}. As one can see, these two
results are very similar to each other, indicating that the group-finding
algorithm cannot bias this relation.

\section{Summary}%
\label{sec:summary}

It is generally believed that the star formation activities of galaxies are
regulated by the formation histories of their host halos. However, this
correlation is difficult to be verified observationally for two reasons. First,
the correlation between galaxy properties and the assembly histories of dark
matter halos is a secondary effect, while the primary correlation is the
stellar mass-halo mass relation. Second, the halo formation histories are not
observable, and finding a model-independent proxy is not trivial. In this
paper, we devised three different models, from the physical hydrodynamical
simulation to the empirical statistical model, to demonstrate the robustness of
using the central stellar mass to halo mass ratio, i.e. central SMHMR, as a
proxy of halo formation time. Then, we made use of the large galaxy sample from
SDSS MGS, together with the physical quantities from the GSWLC catalog and the
group catalog constructed by the halo-based group finder, to infer the relation
between the central galaxy properties and the formation time of their host
halos. Our main results are summarized as follows:

\begin{enumerate}

    \item We established the relation between central SMHMRs and halo formation
        time, where early-formed halos have higher central SMHMRs than their
        late-formed counterparts. This relation is tested on three models, from
        the physical hydrodynamical simulation to the empirical statistical
        model. All three models produce similar scaling relations, supporting
        the robustness of the central SMHMR as a halo formation time proxy.
        (see Fig.~\ref{fig:figures/smhmr_dtform_relation_tng300_am} and
        \ref{fig:figures/dtform_smhmr_relation_tng300_am}).

    \item Using the galaxy group catalog constructed on the SDSS MGS sample, we
        inferred the dependence of the central galaxy properties on the central
        SMHMRs. And we found that halos with high SMHMRs have a higher fraction
        of star-forming/spiral central galaxies than their counterparts with
        low SMHMRs (see Fig.~\ref{fig:figures/ssfr_wrt_csmr_hm_bins_sdss}).

    \item We derived the dependence of the quiescent fraction and the spiral
        fraction on the formation time of their host halos. We found that
        central galaxies in late-formed halos have higher fractions of
        quiescence and lower fractions of spiral morphology by $\lesssim 8\%$
        than their early-formed counterparts (see
        Fig.~\ref{fig:figures/ssfr_wrt_dtform_hm_bins_sdss_2}).

    \item By applying the halo-based group-finding algorithm to a mock catalog,
        we found that it does not introduce any obvious bias into the relation
        between the quiescent fraction of central galaxies and the central
        SMHMRs (see Fig.~\ref{fig:figures/mock_group_finder_result}).

\end{enumerate}

Our results provide the first quantification of the relationship between central
galaxy properties and the formation time of their halos. This relationship can
be further tested using alternative halo formation time proxies, like the
stellar mass gap and the spatial distribution of satellite galaxies
\citep{dariushMassAssemblyGalaxy2010, farahiAgingHaloesImplications2020,
golden-marxObservedEvolutionStellar2022, wechslerConcentrationsDarkHalos2002}.
Meanwhile, these results can be compared with state-of-the-art hydrodynamical
cosmological simulations, like IllustrisTNG
\citep{pillepichSimulatingGalaxyFormation2018,
nelsonIllustrisTNGSimulationsPublic2019}, EAGLE
\citep{schayeEAGLEProjectSimulating2015}, and SIMBA
\citep{daveSIMBACosmologicalSimulations2019}, and test their different
implementations of subgrid physics, which will be presented in the subsequent
paper.

\section*{Acknowledgements}

KW sincerely thanks Prof. Houjun Mo and Prof. Cheng Li for their helpful
discussions. This work is supported by the National Science Foundation of China
(NSFC) Grant No. 12125301, 12192220, 12192222, and the science research grants
from the China Manned Space Project with NO. CMS-CSST-2021- A07.

The authors acknowledge the Tsinghua Astrophysics High-Performance Computing
platform at Tsinghua University for providing computational and data storage
resources that have contributed to the research results reported within this
paper.

Funding for the SDSS and SDSS-II has been provided by the Alfred P. Sloan
Foundation, the Participating Institutions, the National Science Foundation,
the U.S. Department of Energy, the National Aeronautics and Space
Administration, the Japanese Monbukagakusho, the Max Planck Society, and the
Higher Education Funding Council for England. The SDSS Web Site is
http://www.sdss.org/.

The SDSS is managed by the Astrophysical Research Consortium for the
Participating Institutions. The Participating Institutions are the American
Museum of Natural History, Astrophysical Institute Potsdam, University of
Basel, University of Cambridge, Case Western Reserve University, University of
Chicago, Drexel University, Fermilab, the Institute for Advanced Study, the
Japan Participation Group, Johns Hopkins University, the Joint Institute for
Nuclear Astrophysics, the Kavli Institute for Particle Astrophysics and
Cosmology, the Korean Scientist Group, the Chinese Academy of Sciences
(LAMOST), Los Alamos National Laboratory, the Max-Planck-Institute for
Astronomy (MPIA), the Max-Planck-Institute for Astrophysics (MPA), New Mexico
State University, Ohio State University, University of Pittsburgh, University
of Portsmouth, Princeton University, the United States Naval Observatory, and
the University of Washington.

\section*{Data availability}

The data underlying this article will be shared on reasonable request to the
corresponding author. The computation in this work is supported by the HPC
toolkit \specialname[hipp] at \url{https://github.com/ChenYangyao/hipp}.

\bibliographystyle{mnras}
\bibliography{bibtex.bib}

\appendix

\section{The central SMHMR-halo formation time relation in the EAGLE simulation}%
\label{sec:eagle}

\begin{figure*}
    \centering
    \includegraphics[width=0.9\linewidth]{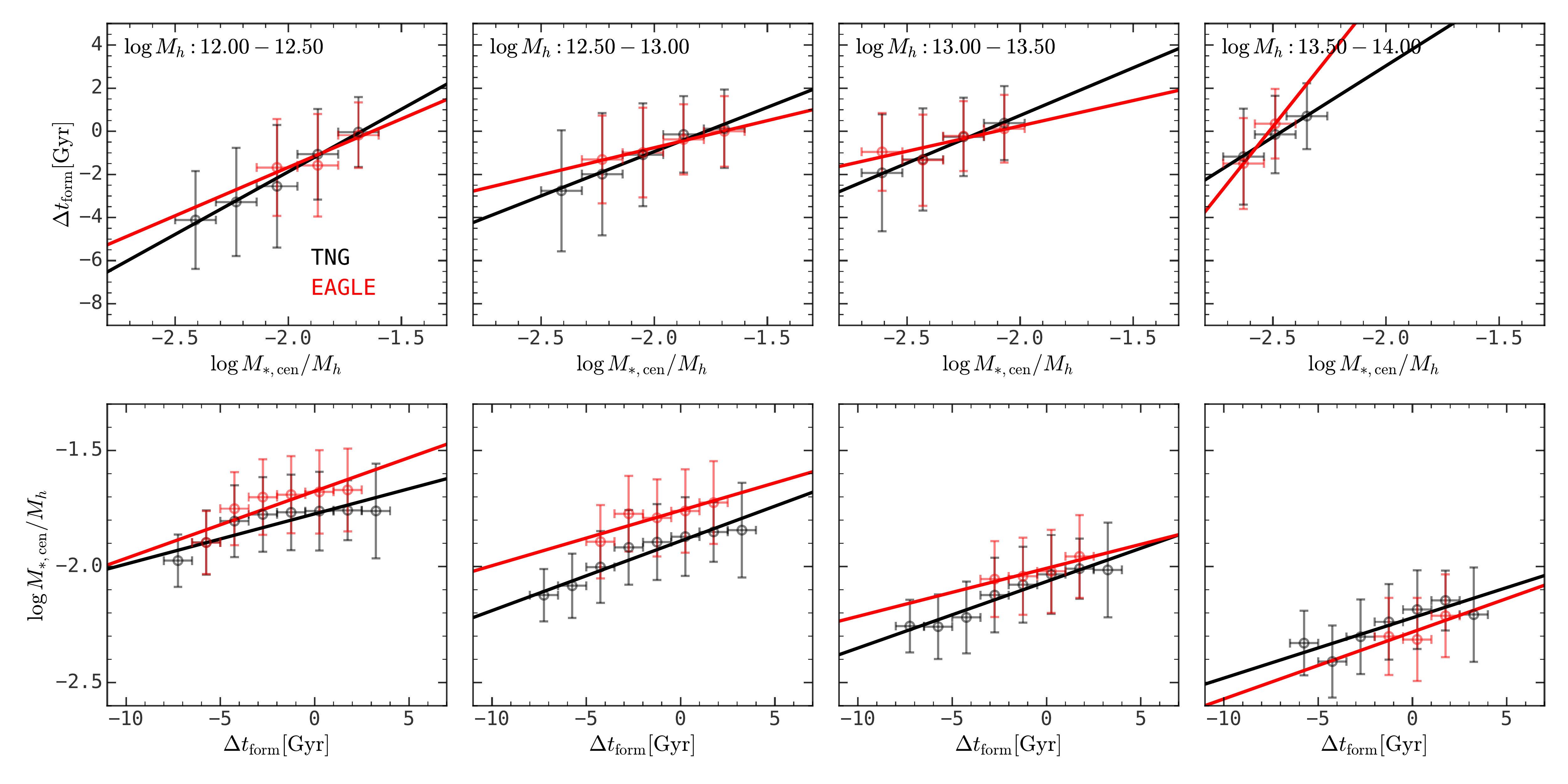}
    \caption{
        The central SMHMR-halo formation time relation for TNG (black) and
        EAGLE (red), where the halo mass is recalibrated using the abundance
        matching method according to the rank of the total stellar mass in each
        main halo.
    }%
    \label{fig:figures/test_with_eagle}
\end{figure*}

Here we show the central SMHMR-halo formation time relation in the EAGLE
simulation \citep{schayeEAGLEProjectSimulating2015}. In this paper, we use the
simulation identified as \texttt{Ref-L0100N1504}, which contains $2\times
1504^3$ particles in a $\rm (100cMpc)^3$ box. The gas particle mass and the
stellar particle mass are $1.81\times 10^6M_{\odot}$ and $9.70\times
10^6M_{\odot}$, respectively. The simulation adopted a flat $\Lambda$CDM
cosmology from the {\it Planck} mission
\citep{collaborationPlanck2013Results2014}, and the cosmological parameters are
$\rm \Omega_m = 0.307$, $\Omega_{\Lambda} = 0.693$, $\rm \Omega_b = 0.04825$,
$\rm\sigma_8=0.8288$, and $h=0.6777$. The halo formation time is defined in the
same way as in the TNG simulation. The stellar mass we use is the sum of all
stellar particles within 30 physical kpc.

Fig.~\ref{fig:figures/test_with_eagle} shows the relation between central
SMHMRs and halo formation time for TNG and EAGLE. One can see that both models
predict that early-formed halos have higher central SMHMRs than their
late-formed counterparts and vice versa. The discrepancy between these two
models is due to the difference in their stellar mass definition and the
implementations of subgrid physics. It is noteworthy to point out that our
results have also been tested on \texttt{Model-C}, which is totally empirical.

\section{Dependence of central galaxy properties on central SMHMR in fine halo mass bins}%
\label{sec:finer_halo_mass_bin}

\begin{figure*}
    \centering
    \includegraphics[width=0.9\linewidth]{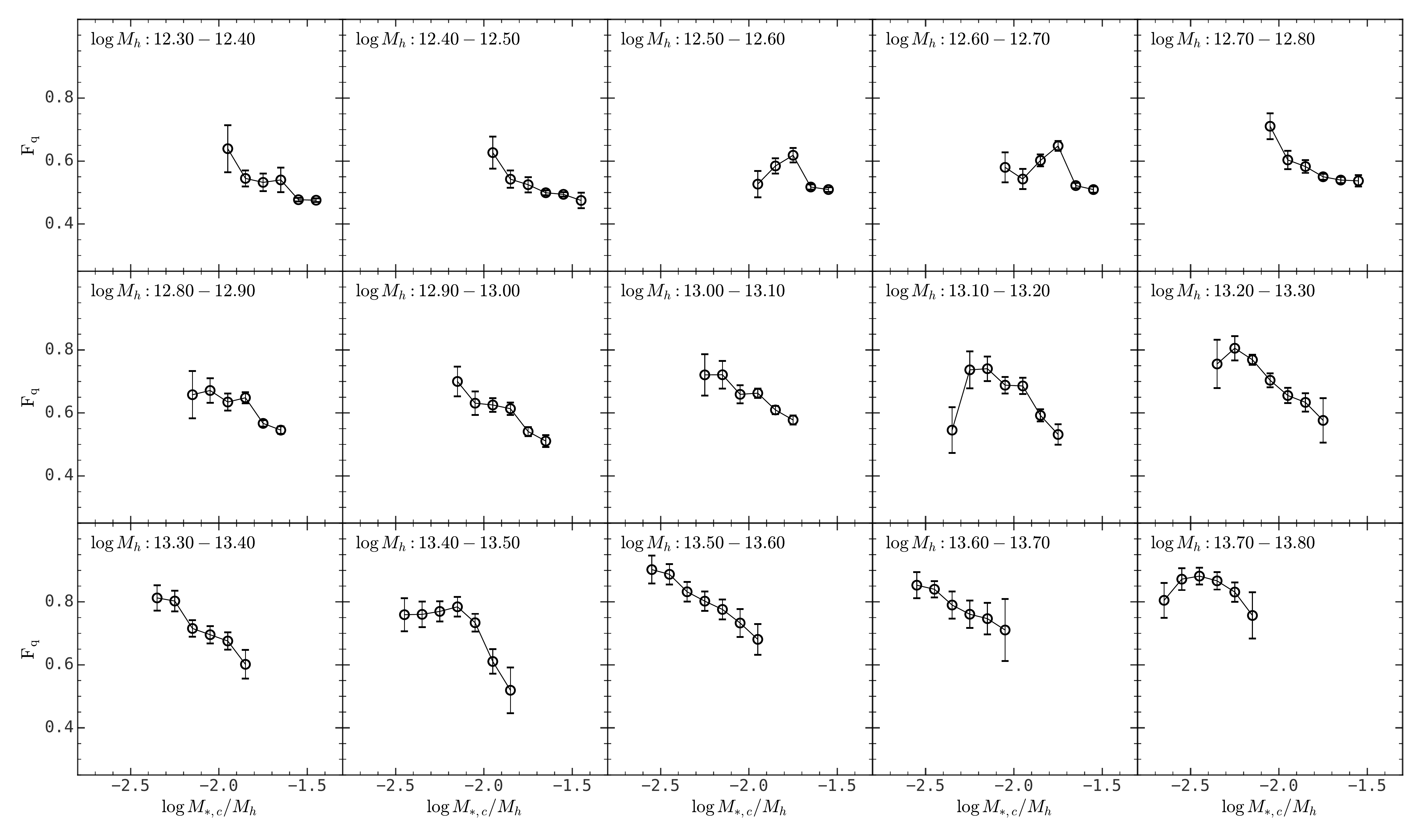}
    \caption{
        Similar to the top panels of
        Fig.~\ref{fig:figures/ssfr_wrt_csmr_hm_bins_sdss}, just in finer halo
        mass bins.
    }%
    \label{fig:figures/ssfr_wrt_csmr_hm_bins_sdss_fine_bin}
\end{figure*}

\begin{figure*}
    \centering
    \includegraphics[width=0.9\linewidth]{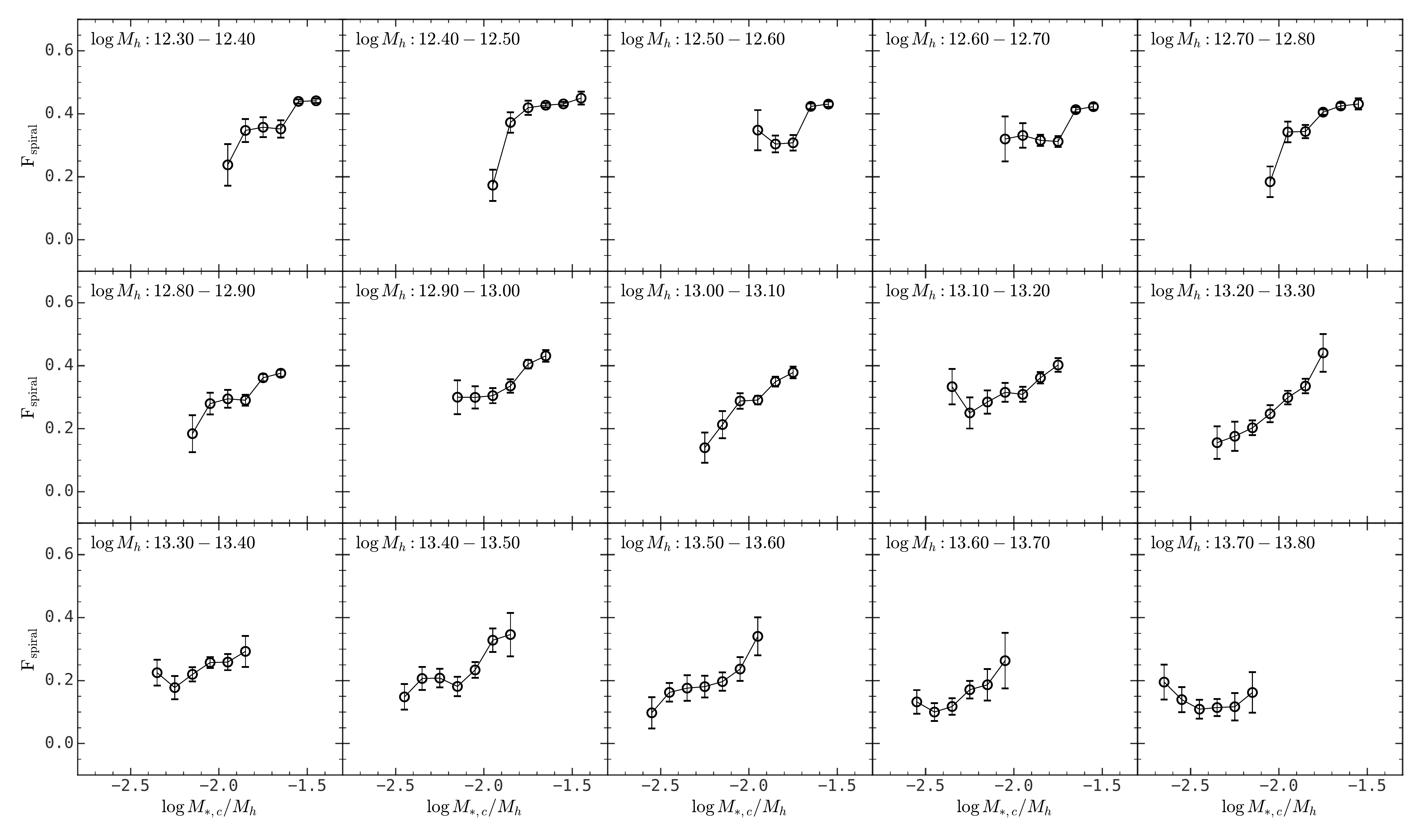}
    \caption{
        Similar to the bottom panels of
        Fig.~\ref{fig:figures/f_spiral_wrt_csmr_hm_bins_sdss_fine_bin}, just in
        finer halo mass bins.
    }%
    \label{fig:figures/f_spiral_wrt_csmr_hm_bins_sdss_fine_bin}
\end{figure*}

The quiescent fraction and the spiral fraction of central galaxies in
Fig.~\ref{fig:figures/ssfr_wrt_csmr_hm_bins_sdss} are in broad halo mass bins
with 0.5 dex width for better statistics. Here
Fig.~\ref{fig:figures/ssfr_wrt_csmr_hm_bins_sdss_fine_bin} and
\ref{fig:figures/f_spiral_wrt_csmr_hm_bins_sdss_fine_bin} show the results in
finer halo mass bins with 0.1 dex width, and one can see that the dependence of
the quiescent fraction and the spiral fraction on central SMHMRs is consistent
with Fig.~\ref{fig:figures/ssfr_wrt_csmr_hm_bins_sdss}.

\section{Dependence of halo formation time on central galaxy properties}%
\label{sec:dependence_tform_on_central}

\begin{figure*}
    \centering
    \includegraphics[width=0.8\linewidth]{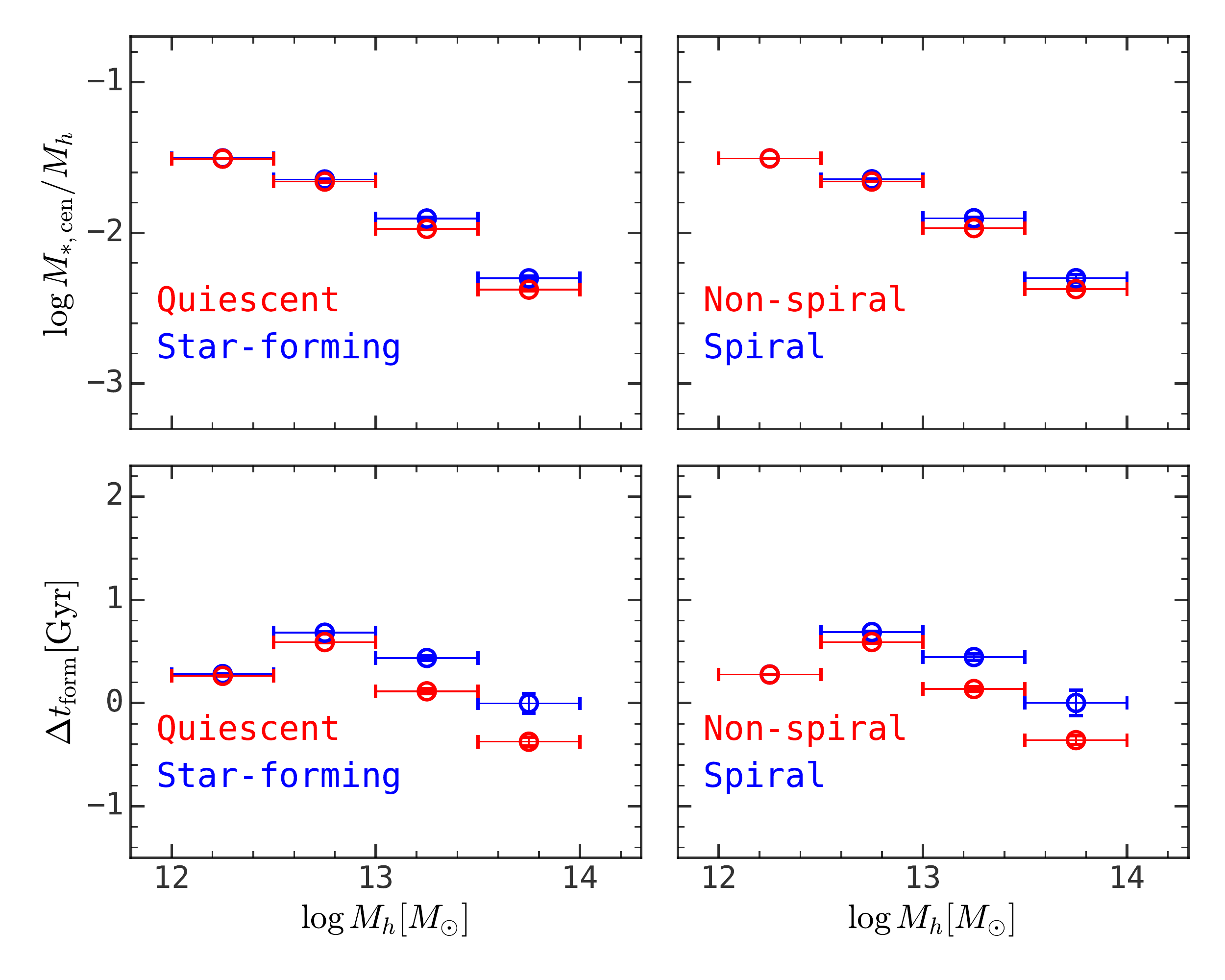}
    \caption{
        {\bf Top panels:} Median central SMHMR as a function of halo mass for
        quiescent/star-forming central galaxies ({\bf left}) and
        spiral/non-spiral central galaxies ({\bf right}). {\bf Bottom panels:}
        Median $\Delta t_{\rm form}$ as a function of halo mass for
        quiescent/star-forming central galaxies ({\bf left}) and
        spiral/non-spiral central galaxies ({\bf right}). All error bars are
        estimated using the bootstrap method.
    }%
    \label{fig:figures/dtform_galaxy_group}
\end{figure*}

\begin{figure}
    \centering
    \includegraphics[width=0.9\linewidth]{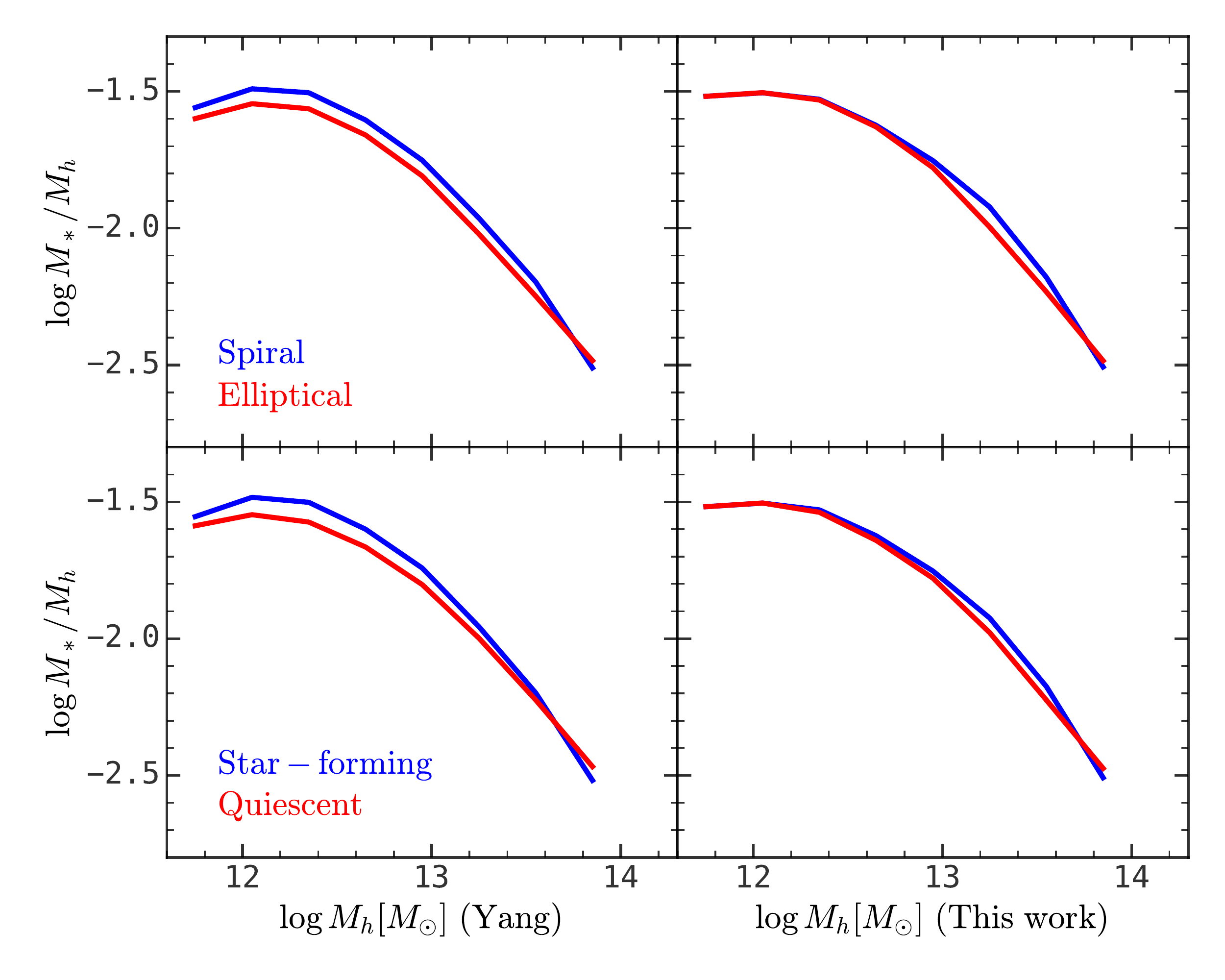}
    \caption{
        Relation between central SMHMRs and halo masses. {\bf Top panels} show
        the results for spiral and elliptical galaxies in blue and red colors,
        respectively. {\bf Bottom panels} show the results for star-forming and
        quiescent central galaxies, respectively. {\bf Left panels} uses the
        halo mass calibrated in \citet{yangGalaxyGroupsSDSS2007} group catalog
        with the rank of the total stellar mass of group members where the stellar
        mass is estimated from $r$-band magnitude and $g-r$ color
        \citep{bellOpticalInfraredProperties2003}. {\bf Right panels} also uses
        the halo mass calibrated according to the rank of total stellar mass,
        but the stellar mass comes from the GSWLC catalog
        \citep{salimDustAttenuationCurves2018}. We note that the stellar mass
        on the $y$-axis is from the GSWLC catalog.
    }%
    \label{fig:figures/impact_of_re_am}
\end{figure}

In \S\,\ref{sub:dependence_smhmr_tform}, we showed that late-formed halos
prefer to host quiescent and non-spiral central galaxies. The correlation
between the halo formation time and the central galaxy properties can also be
demonstrated by examining if the quiescent and non-spiral central galaxies
prefer to live in late-formed halos. Top panels in
Fig.~\ref{fig:figures/dtform_galaxy_group} show the median SMHMR for
quiescent/star-forming and non-spiral/spiral central galaxies with errors
estimated using the bootstrap method. Here one can see that more massive halos
have lower central SMHMR, and the central SMHMR only has a minor dependence on
the properties of central galaxies. Nevertheless, one can still see that
star-forming/spiral central galaxies tend to have higher central SMHMRs than
those quiescent/non-spiral ones. This result is consistent with previous
studies where they found that massive star-forming galaxies have higher central
SMHMR \citep{zhangMassiveStarformingGalaxies2022}, and massive spiral galaxies
also have higher central SMHMR \citep{postiPeakStarFormation2019}.

Here we can convert the central SMHMR to the halo formation time using the
linear fitting functions shown in
Fig.~\ref{fig:figures/smhmr_dtform_relation_tng300_am}, and infer the
difference in the halo formation time for halos with different central galaxy
properties. The results are shown on the bottom panels of
Fig.~\ref{fig:figures/dtform_galaxy_group}, where one can see that massive
halos with star-forming or spiral-like central galaxies are formed early by
$\lesssim 0.4$ Gyr. We note that the halo formation time difference between
star-forming/spiral and quiescent/non-spiral central galaxies is underestimated
here for two reasons. First, from
Fig.~\ref{fig:figures/dtform_smhmr_relation_tng300_am}, one can see that the
correlation between the central SMHMR and the halo formation time is stronger
for \texttt{Model-A}, which means that our heuristic abundance matching method
missed some correlation between these two quantities. Second, due to the
observational limits, some low-mass halos in observation only have one member
galaxy, which produces no difference in their calibrated halo mass.

We note that \citet{correaDependenceGalaxyStellartohalo2020} also inspect the
dependence of central galaxy properties on the central SMHMR, where they find
that blue/disk central galaxies are more massive than their red/elliptical
counterparts for galaxy groups with halo mass less than $10^{13}M_{\odot}$.
However, they find that this signal disappears once a different stellar mass
estimation method is adopted. Here we want to emphasize that this is due to the
inconsistent stellar mass estimators used to calculate central stellar mass and
halo mass. In the group catalog of \citet{yangGalaxyGroupsSDSS2007}, the halo
mass for each galaxy group is estimated by rank-matching the total stellar mass
of galaxy groups and the theoretical halo mass distribution, where the stellar
mass for each galaxy is estimated using the method in
\citet{bellOpticalInfraredProperties2003}. Consequently, central SMHMRs are
equivalent to the central stellar mass to total stellar mass ratios, and if we
use different stellar mass estimators to calculate the central stellar mass and
the total stellar mass, we may introduce artificial biases.

This effect is demonstrated in Fig.~\ref{fig:figures/impact_of_re_am}. On the
left panels, we are showing the central SMHMR as a function of halo mass from
the native galaxy group catalog of \citet{yangGalaxyGroupsSDSS2007}, where the
central stellar mass comes from the GSWLC catalog and the halo mass is from the
total stellar mass estimated in \citet{bellOpticalInfraredProperties2003}. Here
one can see that star-forming/spiral galaxies are more massive than their
quiescent/non-spiral counterparts, which is similar to the result in
\citet{correaDependenceGalaxyStellartohalo2020}. The right panels show the
result where the halo mass is recalibrated according to equation~(\ref{eq:am}),
and both the numerator and the denominator for the central SMHMR are using
consistent stellar mass definition. And one can see that the difference
disappears for halos which mass below $10^{13} M_{\odot}$ and there is still
some difference for halos with mass above $10^{13}M_{\odot}$, as shown in
Fig.~\ref{fig:figures/dtform_galaxy_group}.

% Don't change these lines
\bsp	% typesetting comment
\label{lastpage}
\end{document}